\documentclass[fleqn,usenatbib]{mnras}

\usepackage[T1]{fontenc}
\usepackage{ae,aecompl}
\usepackage{graphicx}
\usepackage{amsmath}
\usepackage{amssymb}
\usepackage{color}

\def\sun{{\odot}}

\title[BH-WD model for GRBs without SN association]
{A Black Hole - White Dwarf Compact Binary Model for Long
Gamma-ray Bursts without Supernova Association}

\author[Dong et al.]{Yi-Ze Dong,
Wei-Min Gu \thanks{E-mail: guwm@xmu.edu.cn},
Tong Liu, and Junfeng Wang \\
Department of Astronomy, Xiamen University, Xiamen, Fujian 361005, China}

\begin{document}
\label{firstpage}
\pagerange{\pageref{firstpage}--\pageref{lastpage}}
\maketitle

\begin{abstract}
Gamma-ray bursts (GRBs) are luminous and violent phenomena in the universe.
Traditionally, long GRBs are expected to be produced by the collapse
of massive stars and associated with supernovae.
However, some low-redshift long GRBs have no detection of
supernova association, such as GRBs 060505, 060614 and 111005A.
It is hard to classify these events convincingly according to
usual classifications, and the lack of the supernova implies
a non-massive star origin.
We propose a new path to produce long GRBs without supernova association,
the unstable and extremely violent accretion in a contact binary
system consisting of a stellar-mass black hole and a white dwarf,
which fills an important gap in compact binary evolution.
\end{abstract}


\section{Introduction}

It is generally believed that long gamma-ray bursts (GRBs) originate from
the collapse of massive stars and are accompanied with supernovae
\citep{Woosley1993ApJ...405..273W,Woosley2006ARA&A..44..507W}.
Traditionally, GRBs 060505, 060614 and 111005A should be classified
as long GRBs because of their long duration time.
However, they are not associated with any
supernova signature even though their redshifts are quite low ($z \la 0.1$).
Such GRBs are also named as long-short GRBs or SN-less long GRBs
\citep[e.g.,][]{Wang2017ApJ...851L..20W}.
Taking GRB 060614 as an example, the temporal lag and peak luminosity
makes it more like a short GRB \citep{Gehrels2006Natur.444.1044G}.
Several models were proposed to explain GRB 060614, including the merger
of a neutron star (NS) and a massive white dwarf (WD) \citep{King2007MNRAS.374L..34K},
and the tidal disruption of a star by an intermediate-mass black hole (BH)
\citep{Lu2008ApJ...684.1330L}.
Recently, a near-infrared bump was discovered in the afterglow of
GRB 060614, which probably arose from a Li-Paczy{\'n}ski macronova,
so the model involved compact binary may be favoured
\citep{Yang2015NatCo...6E7323Y}.

\begin{figure*}
 \begin{center}
  \includegraphics[width=5.4in]{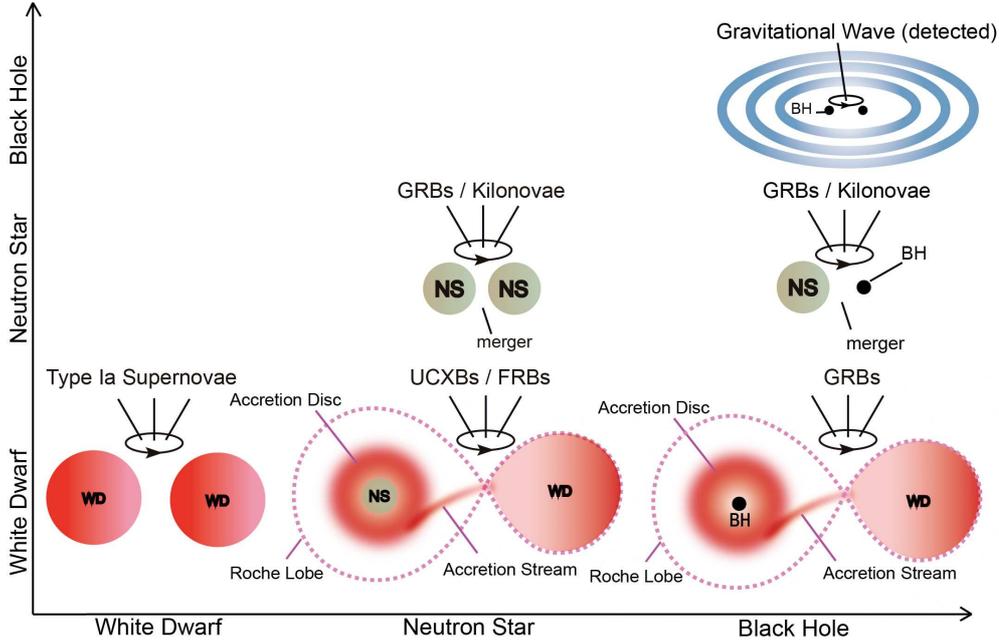}
   \caption{A full picture of compact binary systems. The figure shows
   different compact binary systems and potential corresponding observations.}
   \label{fig:1}
\end{center}
\end{figure*}

Compact binary systems have been widely applied
to different interesting phenomena.
Double WD mergers may produce Type Ia supernovae if their total
mass is larger than the Chandrasekhar limit \citep{Maoz2014ARA&A..52..107M}.
The merger of binary NS, or that of a BH and an NS
is expected to trigger short GRBs and produce Li-Paczy{\'n}ski macronovae
\citep{Li1998ApJ...507L..59L,Metzger2010MNRAS.406.2650M}.
NS-WD systems have been used to explain the ultra-compact X-ray binaries
(UCXBs) \citep{Nelemans2010NewAR..54...87N} and the repeating fast radio burst
(FRB 121102) \citep{Gu2016ApJ...823L..28G}.
In addition, the gravitational wave emission originating from the merger of
double BHs was detected by LIGO \citep{Abbott2016aPhRvL.116f1102A,Abbott2016bPhRvL.116x1103A}.
In this work, we propose that the unstable accretion in a BH-WD system can
explain the long GRBs without supernova association.
With this evolutionary scenario, a full picture of the compact binary systems
may be pieced together and is presented in Figure~\ref{fig:1}.

The gravitational wave emission plays a significant role during the binary evolution.
It carries away energy and angular
momentum and leads to the shrink of the orbit. In addition,
other mechanisms may also remove the orbital angular momentum.
For double degenerate systems, the orbital angular momentum can be converted
to the angular momentum of the accretion disc or to the spin angular momentum
of the accreting star \citep{Ruderman1983Natur.304..425R,Hut1984ApJ...284..675H}.
Our calculations show that there exist strong outflows in the typical
BH-WD system and the orbital angular momentum is carried away, which is similar to those studies on the double
WD system\citep[e.g.][]{Han1999A&A...349L..17H}.
If a typical WD mass $M_{\rm WD}=0.6M_{\sun}$ is chosen for analyses
\citep{Bergeron1992ApJ...394..228B,Kepler2007MNRAS.375.1315K}, the accretion is extremely super-Eddington,
by a factor of $>10^3$ for a stellar-mass BH.
Recent simulations and observations have shown
that, outflows are significant in the super-Eddington case
\citep{Sadowski2015MNRAS.453.3213S,Pinto2016Natur.533...64P,Walton2016ApJ...826L..26W,Fiacconi2017MNRAS.469L..99F,Jiang2017arXiv170902845J}.
Thus, in such a system, outflows can play an important role.

We use an analytic method to investigate the dynamical stability of
the BH-WD system during mass transfer.
For a typical WD donor, when the mass transfer occurs,
the accretion rate can be extremely
super-Eddington and the accreted materials will be ejected from the system.
Such strong outflows can carry away significant orbital
angular momentum and trigger the unstable accretion of the system.
The released energy from the accretion is sufficient to power GRBs.

\section{Model and analyses}

In our model, the system consists of a stellar-mass BH and a WD donor.
In such a system, the mass transfer occurs when the WD fills the Roche lobe.
If this process is dynamically
unstable, the WD will be disrupted rapidly.
The instability of the system greatly depends on the relative expansion
rate of the WD and the Roche lobe. Following the mass transfer,
the radius of the WD will expand, and the orbital separation also tends
to increase if the orbital angular momentum is conserved.
If the WD expands more rapidly than the Roche lobe,
i.e., the timescale for the expansion of the Roche lobe is longer than that of the WD,
the mass transfer will be dynamically unstable.
In this scenario,
the orbital angular momentum loss can restrain the expansion of
the Roche lobe, and therefore will have essential influence on
the binary evolution.

%Now we make detailed analyses of the instability of the system.
We consider a compact binary system which contains a WD donor.
The system is assumed to be tidally locked, and the orbit is circular.
The orbital angular momentum $J$ of the binary system is given by
\begin{eqnarray}
J=M_{1}M_{2}(\frac{Ga}{M})^\frac{1}{2} \ ,
\end{eqnarray}
where $M_1$ and $M_2$ are the mass of the accretor and the WD,
respectively, $a$ is the orbital separation, and $M=M_1+M_2$ represents
the total mass of two stars. A relatively accurate formula for the radius
of WD is provided by Eggleton and is quoted by \cite{Verbunt1988ApJ...332..193V}:
\begin{eqnarray}
\begin{split}
\frac{R_{\rm WD}}{R_\odot}=0.0114\left[(\frac{M_2}{M_{\rm Ch}})
^{-\frac{2}{3}}-(\frac{M_2}{M_{\rm Ch}})^\frac{2}{3}\right]
^\frac{1}{2}\\
\times\left[1+3.5(\frac{M_2}{M_{\rm p}})
^{-\frac{2}{3}}+(\frac{M_2}{M_{\rm p}})^{-1}\right]^{-\frac{2}{3}} \ ,
\end{split}
\end{eqnarray}
where $M_{\rm Ch}=1.44M_\sun$ and $M_{\rm p}=0.00057 M_\sun$.
This relation offers a single relation that can be used in any mass
range of WD \citep{Marsh2004MNRAS.350..113M}. For simplicity, we adopt a form
to describe the Roche-lobe radius of the secondary
$M_2$ (WD)\citep{Paczynski1971ARA&A...9..183P}:
\begin{eqnarray}
\frac{R_{2}}{a}=0.462(\frac{M_{2}}{M})^\frac{1}{3} \ .
\end{eqnarray}
For the circular orbit, the variation of $a$ due to the gravitational
wave emission can be written as \citep{Peters1964PhRv..136.1224P}
\begin{eqnarray}
\frac{da}{dt} = - \frac{64}{5} \frac{G^{3} M_1 M_2 M}{c^5 a^{3}} \ .
\end{eqnarray}
Based on Equations (1)-(4), we can derive the mass accretion rate
for the system,
$$
\dot{M_2}=\frac{64G^3M_1M_2^2M}{5c^5a^4(2q-\frac{5}{3}+\frac{\delta}{3})}
\ ,
$$
$$\delta=\frac{(\frac{M_2}{M_{\rm Ch}})^{-\frac{2}{3}}
+(\frac{M_2}{M_{\rm Ch}})^\frac{2}{3}}
{(\frac{M_2}{M_{\rm Ch}})^{-\frac{2}{3}}-(\frac{M_2}{M_{\rm Ch}})
^\frac{2}{3}}-\frac{2\frac{M_{\rm p}}{M_2}[\frac{7}{3}
(\frac{M_{\rm p}}{M_2})^{-\frac{1}{3}}+1]}
{1+3.5(\frac{M_2}{M_{\rm p}})^{-\frac{2}{3}}+(\frac{M_2}{M_{\rm p}})^{-1}} \ ,
$$
where $q=M_2/M_1$.

\begin{figure}
	\includegraphics[width=\columnwidth]{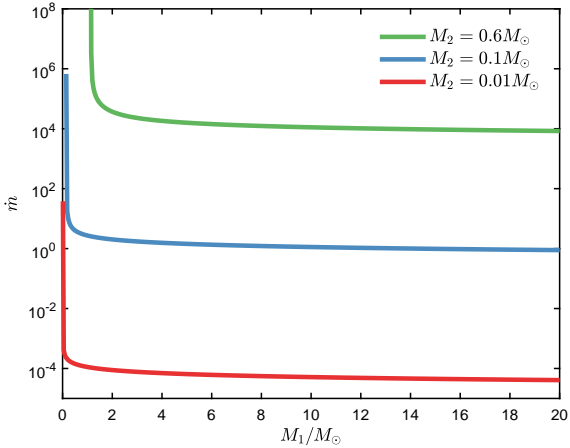}
    \caption{Accretion rates for the BH-WD system with mass transfer.
    The figure shows that the accretion rate is extremely super-Eddington
by a factor larger than $10^4$ for a 0.6$M_{\sun}$ WD.
The critical mass of WD between the sub-Eddington and
super-Eddington cases is around $0.1 M_{\sun}$.}
    \label{fig:3}
\end{figure}

Then, we define a dimensionless parameter $\dot {m}=-\dot{M_2}/
\dot{M}_{\rm Edd}$, where $\dot{M}_{\rm Edd}=L_{\rm Edd}/{\eta c^2}$
is the Eddington accretion rate,
where $L_{\rm Edd}=4\pi GM_1 c / \kappa_{\rm es}$.
We choose $\eta=0.1$. Thus,
$$
\dot m =
\frac{4 G^2 M_2^2 M \kappa_{\rm es}}{25\pi c^4 a^4 (5/6-q-\delta/6)} \ ,
$$
where the opacity $\kappa_{\rm es} = 0.34~{\rm cm}^2~{\rm g}^{-1}$.
For $M_2 = 0.01 M_{\sun}, 0.1 M_{\odot}, 0.6 M_{\sun}$, the variation
of the stable accretion rate with $M_1$ is plotted in Figure \ref{fig:3}.
There exists dramatic increase for $M_1 < 2M_{\sun}$, which indicates that,
for relatively large mass ratio, the mass transfer is unstable for the
system consisting of a WD and an NS/WD.
If $M_1$ is fixed, $\dot{m}$ is positively related to $M_2$.
The critical mass of WD
between sub-Eddington and super-Eddington is around 0.1~$M_{\sun}$.
For $M_2=0.6M_\sun$, the accretion rate is extremely super-Eddington,
so the outflows ought to be taken into consideration.
Here we only considered the effects of gravitational wave emission.
Obviously, outflows can be another mechanism to take away the
angular momentum, so the mass accretion rate should be even higher,
which can result in stronger outflows.
We use a parameter $f$ to describe the scale of mass loss in the mass
transfer process due to outflows:
\begin{eqnarray}
dM_1 + dM_2 = f dM_2\ (0 \leqslant f < 1) \ ,
\end{eqnarray}
where $dM_2$ is negative. In general, the stream of matter flowing from
the inner Lagrange point can form a disc around the accreting star.
If the disc is circular and nonviscous, its radius can be written as
\citep{Verbunt1988ApJ...332..193V}
\begin{eqnarray}
\begin{split}
\frac{R_{\rm h}}{a} = 0.0883 - 0.04858\log q
+0.11489\log^2 q \\ +0.020475\log^3 q \ ,
\end{split}
\end{eqnarray}
where $10^{-3}<q<1$. A fraction of the orbital
angular momentum of the stream comes from the spin angular momentum
of the WD, so this part should be eliminated. Thus, the orbital angular
momentum carried away by the outflows can be expressed as
\begin{eqnarray}
dJ = - \lambda [(d M_{2}\Omega_1R_{\rm h}^2-d M_2\Omega(a-b_1)^2] \ ,
\end{eqnarray}
where $\Omega$ is the orbital angular velocity, $\Omega_1$ is the
orbital angular velocity of the disc, $b_1$ is the distance between
the accreting star and the $L_1$ point, and $\lambda$ is a parameter
probably in the range $0\leqslant \lambda < 1$. The parameter $\lambda$
characterises the loss of orbital angular momentum through outflows.
The expression of $b_1$ takes the form:
\begin{eqnarray}
\frac{b_1}{a}=0.5-0.227 \log q \ .
\end{eqnarray}

%When the WD fills its Roche lobe, mass loss from the WD starts.
%Following the mass transfer, the radius of WD expands and the Roche
%lobe also changes.
The instable mass transfer will occur if the radius of WD expands
more rapidly than the Roche lobe, i.e.,
$\dot{R}_{\rm WD}/R_{\rm WD}>\dot{R_2}/R_2$.
Based on Equations (1)-(3) and (5)-(8), we can derive the instability
criterion for the mass transfer:
\begin{eqnarray}
\begin{split}
{\mathcal F} (q,f,\lambda) = \lambda (1+q) \left( \frac{b_1-a_1}{a} \right)^2\\
-\frac{5-\delta}{6} + q(1-f) + \frac{fq}{3(1+q)} > 0 \ ,
\label{eq:criterion}
\end{split}
\end{eqnarray}
where $a_1$ is the distance
between the BH and the mass centre of the system.
We adopt $f=0.9$ and $f=0.99$ for calculations, which means that 90\% or 99\%
materials transferred from the WD are lost the system due to strong outflows.
The material stream ejected from the system may
form a common envelope around the binary system \citep{Han1999A&A...349L..17H}.
We assume $\lambda \geqslant f$ since the orbital angular momentum can be
carried away by the outflows. Moreover, the outflows may possess larger
angular momentum per unit mass than the inflows thus can escape from
the system.

\begin{figure}
	\includegraphics[width=\columnwidth]{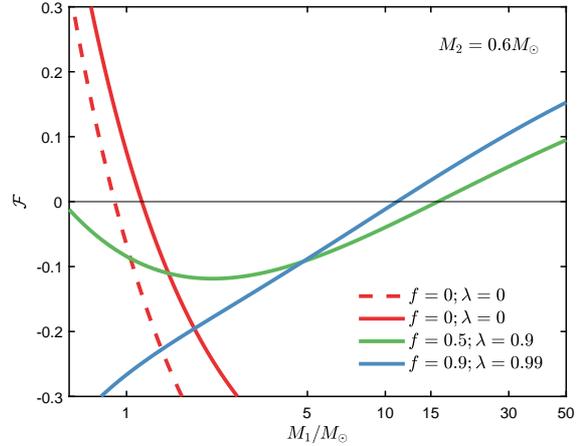}
    \caption{Criterion for an unstable mass transfer.
    The red dashed and solid lines represent the results of \citet{King2007MNRAS.374L..34K}
and ours, respectively, where outflows are not considered.
The green and blue lines correspond to our results including the
effects of outflows. The criterion for an unstable mass transfer
is $\mathcal{F} >0$, as shown by Equation~(\ref{eq:criterion}).}
    \label{fig:4}
\end{figure}

\section{Results}
\label{S:Results}

The results for $M_2 = 0.6 M_\sun$
are shown in Figure~\ref{fig:4}. An unstable
region exists for $M_1 > 15.97 M_\sun$ under $f=0.5$ and $\lambda =0.9$
(green line), and for $M_1 > 11.09 M_\sun$ under $f=0.9$
and $\lambda =0.99$ (blue line).
The required mass of $M_1$ for the unstable accretion
corresponds to a BH system.
For comparison, the results under the conservative condition,
i.e., $f=0$ and $\lambda=0$, is plotted by the red solid line,
and the critical mass ratio is $q=0.52$.
Such a result is similar to previous studies
\citep{Ruderman1983Natur.304..425R,Hut1984ApJ...284..675H,Verbunt1988ApJ...332..193V}.
In this scenario, only when the accretor mass is smaller than
about 1.15 $M_\sun$, the accretion can be unstable, which means that
the accretor cannot be a BH.
The results of \citet{King2007MNRAS.374L..34K} are slightly different from ours
(red dashed line), since an elaborated formula for the radius of WD
is adopted in the present work.

\begin{figure}
	\includegraphics[width=\columnwidth]{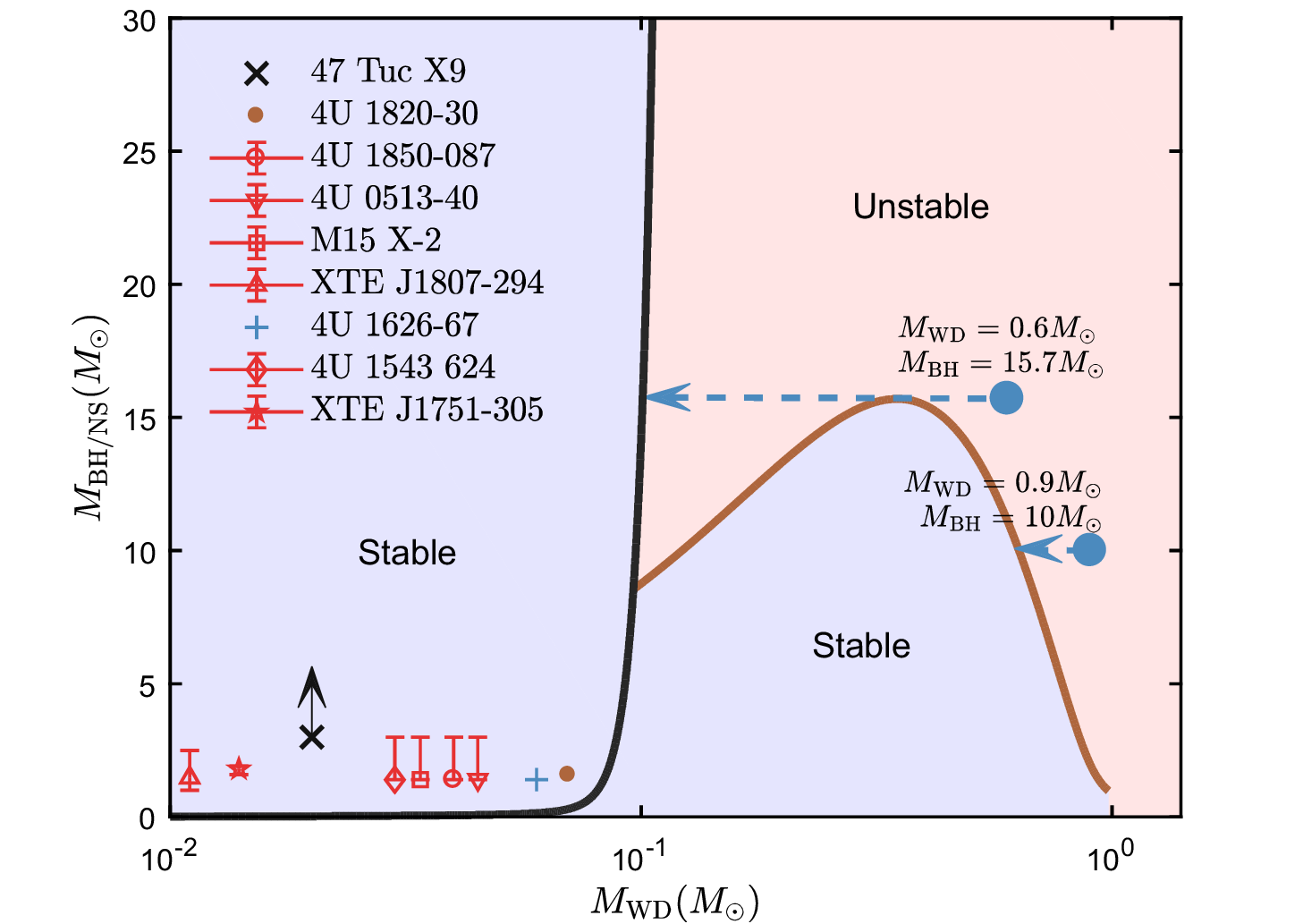}
    \caption{Variation of the critical BH mass with the WD mass for $f=0.9$
and $\lambda=0.99$. For the parameters beyond the critical BH mass (brown line),
the mass transfer is dynamically unstable.
The two blue dashed lines show the evolutionary track
(from the filled circle to the arrow).
The space is divided into two regions, sub-Eddington (left)
and super-Eddington (right), by the black solid line.
The nine UCXBs are well located in the sub-Eddington and stable region,
where eight UCXBs are NS candidates and the other one is a BH candidate.}
    \label{fig:2}
\end{figure}

The strong outflows during mass transfer can carry away angular momentum,
which influences the stability of the system.
In reality, the mass of the
components and the separation of the binary also
change due to the mass transfer during the evolution,
so the system may not always be unstable.
We plot the critical condition (brown line)
in Figure \ref{fig:2}, where $f=0.9$ and $\lambda=0.99$ are adopted for
the super-Eddington accretion case.
It should be noted that once the WD fills its Roche lobe within
the unstable region, the violent accretion will continue
as long as the evolutionary track is still located in the unstable region.
In Figure \ref{fig:2}, the brown line has a maximum value
$M_{\rm BH} = 15.7 M_{\sun}$ at $M_{\rm WD}= 0.35 M_{\sun}$.
Thus, for a typical WD with $M_{\rm WD}= 0.6 M_{\sun}$,
the system can maintain the unstable mass transfer if the BH is
sufficiently large $M_{\rm BH} > 15.7 M_{\sun}$, i.e.,
the accretion process will not stop until the accretion
becomes sub-Eddington, corresponding to a low-mass WD around
$0.1 M_{\sun}$ (the nearly vertical gray line in Figure~\ref{fig:2}).
The blue dashed line shows the mass changes of components along with the mass transfer.
If the BH mass is lower, less material of the WD can be accreted by the BH.
For example, for $M_{\rm WD}= 0.9 M_{\sun}$ and $M_{\rm BH} = 10 M_{\sun}$,
the critical mass is $M_{\rm WD}= 0.63 M_{\sun}$
for $M_{\rm BH} = 10 M_{\sun}$, and therefore
only around $0.27 M_{\sun}$ will be accreted under the unstable phase.
If we consider a BH with larger mass, such as 20 $M_{\sun}$,
for a 0.6~$M_{\sun}$ WD, the accreted material is around 0.5~$M_{\sun}$.
Recently, the first BH UCXB candidate X9 has been discovered
in globular cluster 47 Tucanae
\citep{Bahramian2017MNRAS.467.2199B,Miller2015MNRAS.453.3918M},
and its highest accretion rate is estimated as a few times of
$10^{-10} M_{\sun}~\rm yr^{-1}$, which suggests that the BH mass can
reach a few dozens of solar mass. In this system, however, the WD mass
is only around 0.02$M_{\sun}$, and therefore such a system is under the
stable type of mass transfer, as shown in Figure~\ref{fig:2}.
Nine UCXB candidates are plotted, including eight NS UCXB candidates,
4U 1626$-$67 \citep{Takagi2017nuco.confb0301T},
4U 1850$-$087, 4U 0513$-$40, M15 X$-$2
\citep{Prodan2015ApJ...798..117P},
XTE J1751-305 \citep{Andersson2014MNRAS.442.1786A,Gierlinski2005MNRAS.359.1261G},
XTE J1807$-$294 \citep{Leahy2011ApJ...742...17L},
4U 1820$-$30 \citep{Guver2010ApJ...719.1807G}
and 4U 1543$-$624 \citep{Wang2004ApJ...616L.139W},
and a BH UCXB candidate, 47 Tuc X9.
It is shown that all the UCXB candidates are well located in the left
``Stable" region. Such a location is quite reasonable since the unstable
mass transfer corresponds to extremely high accretion rates and therefore
short timescale for existence, whereas the left stable region corresponds
to sub-Eddington rates and therefore long timescale.

The isotropic energy for a GRB is around $10^{52}$~erg, and can be
up to $10^{54}$~erg.
It is difficult to calculate the accurate energy released from the
unstable accretion of
the BH-WD system. Thus, a rough estimate of the total released energy
can be expressed as:
\begin{eqnarray}
E_{\rm tot}=M_{\rm a}(1-f)\eta c^2 \ ,
\end{eqnarray}
where $M_a$ is the material disrupted by BH. For $M_{\rm WD} = 0.6 M_{\sun}$, $M_{\rm BH}=15.7 M_{\sun}$, $f=0.9$,
and $\lambda=0.99$, $M_{\rm a}=0.25 M_{\sun}$. We adopt $\eta = 0.1$,
so the total energy $E_{\rm tot}$ is around $10^{52}$~erg.
For a typical half-opening angle $\Delta\theta =
10^{\circ}$ for GRBs, the isotropic energy is $E_{\rm tot}/(1-\cos\theta)
\approx 6.6\times 10^{53}$erg, which is sufficient to power most GRBs.
Obviously, it is impossible to detect a supernova in such a event.
The current merger rate for BH-WD binary is about $1.9 \times 10^{-6}
\rm ~yr^{-1}$ in the galactic disc \citep{Nelemans2001A&A...375..890N},
with typical BH mass around $5\sim 7~M_{\sun}$,
which is slightly less than the typical required BH mass
in our model, so the event rate for our system is lower.
For a BH-WD binary system, the initial system should contain
two main-sequence stars, a large mass primary star ($> 25M_{\sun}$) and
a secondary star ($1\sim 8M_{\sun}$). The primary star will evolve into
a $3\sim 15~M_{\sun}$ BH via a supernova explosion
\citep{Woosley1995ApJS..101..181W,zhang2008ApJ...679..639Z},
whereas the secondary star still stay at the main sequence.
The system will experience a common envelope phase when
the secondary star evolves off the main sequence, and finally become a WD.
The initial separation of the compact system should be
comparable to the separation of the X-ray binary,
so we can roughly estimate the time required for
the binary system to evolve into contact according to equation (4).
For an X-ray binary, the secondary star fills its Roche lobe,
so the separation of the X-ray binary can be estimated by Equation (3),
while the radius of the main sequence star is given by
\citet{Demircan1991Ap&SS.181..313D}:
$R_{\rm m} = 1.01 M_{\rm m}^{0.724} (0.1 M_{\sun} < M < 18.1 M_{\sun})$,
where $R_{\rm m}$ and $M_{\rm m}$ are the radius and the mass of
the main sequence star, respectively.
For a system contains a 5-15$M_{\sun}$ BH and 1-8$M_{\sun}$ main sequence
star, after the main sequence star become a WD, the time required for
the WD to fill its Roche lobe
is about $10^9$ year, which is shorter than a Hubble time.
\citet{Bahramian2017MNRAS.467.2199B} 
reported that they detected a contact binary system with a BH and a WD.
The accretion in this system ought to be stable due to the low mass WD.
It implies that more undetected BH-WD systems may exist.
For the unstable mass transfer,
the WD can be disrupted within a few orbital periods \citep{Fryer1999ApJ...520..650F}.
The orbital period can be simply estimated as
$P \approx 46~(M_{\sun}/M_2)~{\rm s}$.
For a typical WD $M_2=0.6M_{\sun}$, we have $P=77$~s. Thus, the accretion
timescale is roughly in agreement with that of the long GRBs.
For GRB 060614, the burst lasts about 100~s,
and such a timescale is reasonable in our model.

\section{Discussion}

We have shown that outflows can play a significant role in the stability
of a contact BH-WD binary system,
and the WD can be nearly disrupted by the BH in a short timescale.
In our binary system, the gamma-ray burst will take place along
the rotational axis, and the outflows will spread over
the orbital plane. Thus, the environment should not be baryon-rich,
and the corresponding spectral signature may be quite weak even if it exists.
It should be noted that the dynamical region for conserved
mass transfer ($q>0.52$) disappears when outflows are considered.
For the extreme super-Eddington case, strong outflows are inevitable.
Thus, our model including the effects of outflows may be more reasonable.
\citet{Fryer1999ApJ...520..650F}
conducted numerical simulations of the mergers of WDs and BHs and proposed
that these mergers can explain long GRBs.
They assumed that no mass is ejected from the system, so the orbital
angular momentum in their simulations is carried away by accretion discs
instead of outflows. The studies about NS-WD or double WD mergers give
a similar critical unstable mass ratio $q < 0.002-0.005$
\citep{Ruderman1983Natur.304..425R,Hut1984ApJ...284..675H,Ruderman1985ApJ...289..244R,Bonsema1985A&A...146L...3B}.
orbital angular momentum conversion mechanisms.
However, in their models, the angular momentum loss
can occur only when the mass ratio is less than the critical value before
the mass transfer starts. Otherwise, the angular momentum can be
transferred back to the orbit through tidal interaction, and the system
turns back to the conserved condition. In our model,
a large fraction of matter is expelled from the system
after the formation of the accretion disc and carries away angular momentum,
so the angular momentum of the system should always be non-conserved
if the mass ratio is less than the critical value.

\section*{Acknowledgements}

We thank Taotao Fang for beneficial discussion, and thank the referee
for helpful suggestions that improved the manuscript.
This work was supported by the National Basic Research Program of China
(973 Program) under grants 2014CB845800,
and the National Natural Science Foundation of China under grants 11573023,
11522323, 11473022, 11473021, and 11333004.

\bsp
\label{lastpage}

\end{document}